\documentclass[amsmath, amssymb, superscriptaddress, showpacs, showkeys, notitlepage, longbibliography, pra, twocolumn]{revtex4-1}

\usepackage{graphicx}
\usepackage[usenames, dvipsnames]{color}
\usepackage{dsfont}
\usepackage[utf8]{inputenc}
\usepackage[english]{babel}

\usepackage[breaklinks=true]{hyperref}
\hypersetup{
  colorlinks   = true, 
  urlcolor     = blue, 
  linkcolor    = blue, 
  citecolor   =  red 
}

\graphicspath{ {./Pictures/},{./MathematicaPlots/} }

\newcommand{\ket}[1]{\ensuremath{\left|#1\right\rangle}}

\newcommand{\Px}{X}	
\newcommand{\Nn}{N_\textrm{S}}	

\newcommand{\hc}{^{\dagger}}							
\newcommand{\HC}{\textrm{h.c.}}
\newcommand{\ee}{\mathrm{e}}						
\newcommand{\ii}{\mathrm{i}}			             			
\renewcommand{\H}[0]{H}  							

\newcommand{\comm}[2]{\left[ #1, #2 \right]} 				
\newcommand{\nn}{\nonumber}							
\newcommand{\abss}[1]{\ensuremath{ \left| #1 \right|^{2} }}	
\newcommand{\diss}[1]{\mathcal{D}[ #1 ]}					
\renewcommand{\l}[0]{\left}
\renewcommand{\r}[0]{\right}

\newcommand{\Tr}{\text{Tr}}

\newcommand{\fig}[1]{Fig.~\ref{#1}}

\begin{document}




\title{Breaking time-reversal symmetry with a superconducting flux capacitor\\
\emph{or}\\
Passive on-chip, superconducting circulator using a ring of tunnel junctions}

\author{Clemens~M\"uller}
\email{clemens.mueller@phys.ethz.ch}
\affiliation{ARC Centre of Excellence for Engineered Quantum Systems, School of Mathematics and Physics, The University of Queensland, Brisbane, QLD 4072, Australia}
\affiliation{Institute for Theoretical Physics, ETH Z\"urich, 8093 Z\"urich, Switzerland}

\author{Shengwei~Guan}
\affiliation{ARC Centre of Excellence for Engineered Quantum Systems, School of Mathematics and Physics, The University of Queensland, Brisbane, QLD 4072, Australia}

\author{Nicolas~Vogt}
\affiliation{Chemical and Quantum Physics, School of Science, RMIT University, Melbourne VIC 3001, Australia}

\author{Jared~H.~Cole}
\affiliation{Chemical and Quantum Physics, School of Science, RMIT University, Melbourne VIC 3001, Australia}

\author{Thomas~M.~Stace}
\affiliation{ARC Centre of Excellence for Engineered Quantum Systems, School of Mathematics and Physics, The University of Queensland, Brisbane, QLD 4072, Australia}

\begin{abstract}
	We present the design of a passive, on-chip microwave circulator based on a ring of superconducting tunnel junctions. 
	We investigate two distinct physical realisations, based on Josephson junctions (JJ) or quantum phase slip elements (QPS), 
	with microwave ports  coupled either capacitively (JJ) or inductively (QPS) to the ring structure.
	A constant bias applied to the center of the ring provides an effective symmetry breaking field, and no microwave or rf bias is required.
	We show that this design offers high isolation, robustness against 
	 fabrication imperfections and bias fluctuations, 
	and has a bandwidth in excess of 500 MHz for realistic device parameters.
\end{abstract}


\date{\today}
\maketitle

Microwave circulators~\cite{Hogan1952, Hogan1953} are ubiquitous microwave circuit elements~\cite{Clarke2008, You:N:2011, Gu:PR:2017}, for signal routing and signal/control isolation. 
{They are also key non-reciprocal elements for realising chiral quantum optics~\cite{Lodahl:N:2016} with microwave photons as well as 
for microwave photon detection~\cite{Sathya:PRL:2014} and rectification~\cite{Muller:2017, Peng:NP:2014}. 
Commercial, passive circulators are wave-interference devices based on the Faraday effect, which require permanent magnets to break time-reversal symmetry. 
Size, and their  strong magnetic fields, make them unsuited to large-scale integration with superconducting circuits, 
presenting a hurdle for scaling-up superconducting quantum technology.}

With the exception of \citet{Koch:PRA:2010}, most recent approaches to this problem use active devices, based on non-linear mixing phenomena~\cite{Kamal2011, Estep2014, Sliwa2015, Lecocq:PRA:2017}
or engineered interplay of driving and dissipation~\cite{Kamal:PRA:2017, Fang:NP:2017, Metelmann:PRX:2015}. 
These proposals rely on careful engineering of phase relations between several input and drive fields. 
Using an rf-driven inductive bridge circuit, \citet{Kerckhoff2015a} demonstrated 
bandwidths  $\sim100$~MHz and tuneable centre frequency~\cite{Chapman:2017, Rosenthal:2017}.   
Passive  unidirectional devices based on Quantum Hall edge modes \cite{stace2004mesoscopic,Viola2014}, 
have been demonstrated~\cite{Mahoney:PRX:2017,Mahoney:2017}.  However there are challenges fabricating these elements in a superconducting circuit. 

\begin{figure}
	\includegraphics[width=\columnwidth]{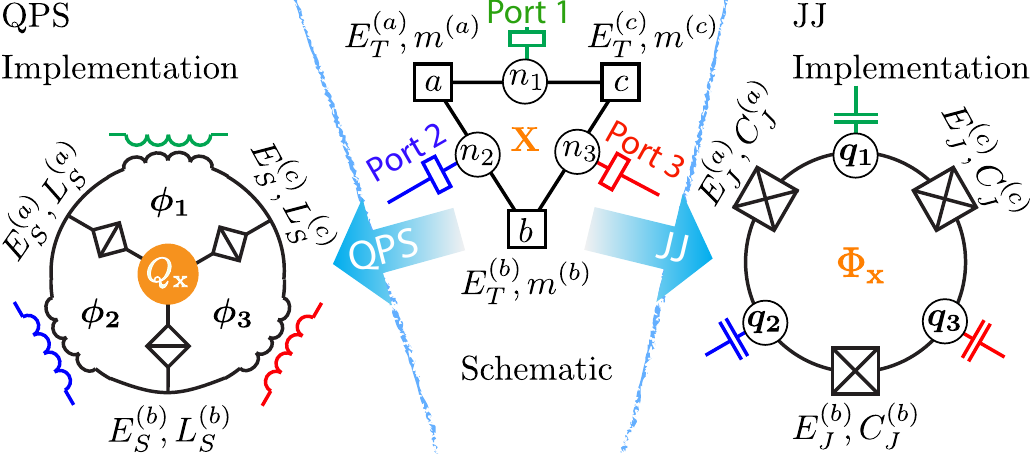}
	\caption{(Color online) 
		(Centre) Schematic representation of the circulator, consisting of three ports connected via coupling elements to the numbered nodes of the ring to the coordinate $n_j$ associated to node $j$.  
		{A central ring bias $\Px$  is conjugate to the $n_j$. Nodes of the ring are mutually coupled by tunnelling elements with tunnelling energy $E_T^{(k)}$ 
		and  `mass'  $m_{T}^{(k)}$.  
		Notionally, the tunnelling elements are identical. Differences lead to imperfect operation.}  
		(Left) The QPS implementation of the scheme using flux tunnelling and capacitive bias. Here $n_j\rightarrow \phi_j/\Phi_0$ are coupled inductively to the external lines, 
		$E_T^{(k)}\rightarrow E_S^{(k)}$ is the phase slip energy, 
		$m_{T}^{(k)}=L_S^{(a)}$ is the QPS inductance, and $\Px \rightarrow Q_\textrm{x}/(2e)$ is the linked charge. 
		(Right) The JJ implementation of the scheme relying on charge tunnelling and inductive bias. Then $n_j \rightarrow q_j/(2e)$ is coupled capacitively to the external lines, 
		$E_T^{(k)} \rightarrow E_J^{(k)}$ is the Josephson energy, $m_{T}^{(k)}=C_J^{(k)}$ is the JJ capacitance, and $\Px\rightarrow\Phi_\textrm{x}/\Phi_0$ is the linked flux. 
	}
	\label{fig:Circuits}
\end{figure}

In this Letter, we provide a detailed theoretical analysis of a fully passive, integrated superconducting microwave circulator realised as a ring of tunnel junctions. 
We simultaneously analyse two implementations of the system, one based on Josephson junctions (JJ), 
which are in common use for quantum information applications~\cite{Barends:N:2016, Langford:2016, Braumuller:2016}, 
and the other based on quantum phase slip (QPS) wires \cite{Mooij:NP:2006}.  
QPS junctions are dual to JJ's  under the exchange of voltage and current~\cite{Mooij:NP:2006}, and they have recently been employed to observe coherent quantum phase slips~\cite{Astafiev:N:2012} 
and as the basic building block of a new type of flux qubit~\cite{Peltonen:PRB:2013, Peltonen:PRB:2016}.  
The underlying mathematical description of these circuit elements is a precise duality, however they have different noise, fabrication and geometric characteristics, 
so that the two implementations may be apposite to different applications or materials. 

The basic physics behind our circulator proposal is the non-local phase accumulation in the Aharanov-Casher effect for  QPS devices, or its dual, the Aharanov-Bohm effect for  JJ devices.   
Both effects arise from the non-local topological mutual phase that charge and flux quanta acquire as they are transported around one another.  

The operation of our proposal is similar to that of \citet{Koch:PRA:2010}, but with a number of significant theoretical and practical differences. 
{It does not require extraneous resonators in the devices, nor any active microwave or rf circuitry, both of which  simplifies and substantially shrinks the circuit. Further,} we calculate scattering matrices in a fully dynamic picture that includes the internal degrees of freedom of the circulator {without relying on an approximate perturbative treatment.}
{Going beyond a linearised, harmonic approximation} enables us to quantify the performance of the device at high coupling energies and with high fluxes, both of which preclude perturbative treatments.   
We show that with experimentally reasonable parameters 
passive, on-chip circulators can be built with bandwidths \mbox{$\sim 500$ MHz}, and with moderate photon flux. 

\paragraph*{Hamiltonian:} To facilitate this dual description, we refer to the diagram  in the centre  of \fig{fig:Circuits}.  
{External ports (numbered $j=1$ to 3) are coupled to `segments' (circles) arranged in a ring, with canonical 	`momenta' $n_j$.
The segments of the ring are mutually coupled through tunnelling elements (squares),  characterised by a tunnelling energy $E_T$ and a `mass' term $m_{T}$.  
The ring of segments encircle a central bias, $\Px$, 
providing a time-reversal-symmetry breaking (effective) magnetic field.} 
 
Physically, in the QPS implementation, the segment degrees of freedom correspond to fluxes threading the spokes of the ring structure, 
illustrated on the left of \fig{fig:Circuits}, i.e.\ $n_j\rightarrow \phi_j /\Phi_{0}$, and the central bias is a charge bias, $\Px\rightarrow Q_\textrm{x} / (2e)$. 
The coupling to external degrees of freedom is realised via a coupling inductance $L_{C}$ with an associated coupling mass term $m_{C}$. 
Additionally, each segment has a parasitic inductance $L_{G}$, corresponding to a final mass term $m_{G}$ in the general description.
Conversely, in the JJ implementation, the segment degrees of freedom correspond to charges at the nodes between two Josephson junctions, 
as illustrated on the right of \fig{fig:Circuits}, i.e.\ $n_j\rightarrow q_j / (2e)$, and the central bias is a flux bias, $\Px\rightarrow  \Phi_\textrm{x} / \Phi_{0}$. 
Coupling to the ring is realised capacitively, $m_{C} = C_{C}$ and each node has an additional parasitic capacitance $m_{G} = C_{G}$. 
In both cases, eigenmodes of the ring  have  flux or charge currents circulating around the ring, which acquire phases dependent on the central bias $\Px$, 
through the Aharanov-Casher/Bohm effect~\cite{Friedman:PRL:2002, Mueller:PRB:2013}.  
Interference between different ring excitations leads to the non-reciprocity required for circulation. 
  
The quantised Hamiltonian for the ring is~\cite{Koch:PRA:2010, Vool:2017}
\begin{align}
	H_{\text{Ring}} =& \tfrac{p_{0}^{2}}{2}( \hat{\mathbf  n }- \mathbf \Nn) \mathds{M}^{-1} (\hat{\mathbf  n }- \mathbf \Nn) \nn\\&
	 - E_{T} {\sum}_{j} \cos{(2\pi (\hat x_{j+1} -\hat x_{j } - \Px/3))},
	\label{eq:HRing}
\end{align}
where $\hat {\mathbf  n} =\{\hat n_1,\hat n_2,\hat n_3\}$ are dimensionless dynamical variables, $\mathbf N_{n}=\{\Nn^{(1)},\Nn^{(2)},\Nn^{(3)}\}$ are the classical bias offsets for each segment,  
$\mathds{M}=m_\Sigma \mathds{1}_{3}-m_{T}$ is the mass tensor and $\hat x_j$ is the conjugate variable to $\hat n_j$, i.e.\ $[\hat n_j,\hat x_j]=i$. 
$m_{\Sigma} = 3 m_{T} + m_{C} + m_{G}$ represents an effective total mass of excitations in the ring and provides the scale of the kinetic energy term. 
Here $p_{0}$ plays the role of zero-point `momentum' in the Hamiltonian and depends on the physical implementation chosen: 
for  QPS devices $p_{0} = \Phi_{0}$ is the superconducting flux quantum; for  JJ devices  $p_{0} = 2e$ is the cooper pair charge. 
Eq.~\eqref{eq:HRing} assumes a rotationally symmetric ring, where all mass and tunnelling energies are equal. 
Generalising to disordered structures simply changes the mass tensor and the tunnelling energies, c.f. supplementary material S.1 and S.5.

We change from local to collective coordinates: $\hat n_{1}' = \hat n_{1}$, $\hat n_{2}' = -\hat n_{2}$, and  $\hat n_{3}' = \hat n_{1} + \hat n_{2} +\hat  n_{3} = N_{0}$.
The latter is the conserved total charge of the ring (since $\partial H / \partial \hat x'_{3} = 0$), so that the Hamiltonian becomes
\begin{align}
	H_{\text{Ring}} =  \frac{p_{0}^{2}}{m_{\Sigma}} \Big(& \bigl( \hat{n}_{1}' - (N_{0} + \Nn^{(1)} - \Nn^{(2)})/2 \bigr)^{2} \nn\\
		+& \bigl( \hat{n}_{2}'+ (N_{0} + \Nn^{(2)} - \Nn^{(3)})/2 \bigr)^{2} - \hat{n}_{1}' \hat{n}_{2}' \Big) \nn\\
		{}- E_{T} \big( &\cos{( 2\pi (x_{1}' - \Px/3))} + \cos{( 2\pi (x_{2}' - \Px/3))} \nn\\
		+& \cos{( 2\pi (x_{1}' + x_{2}' + \Px/3))} \big) \,.
\end{align}
The value of $N_{0}$ is fixed to its ground-state value by the choice of segment bias parameters $\mathbf{N}_{\mathrm S}$~\cite{Koch:PRA:2010}.

The ring couples to external waveguides at each port, described by the  Hamiltonian
\mbox{{$
	H_{WG}={\sum}_{k,j} \omega_k\hat a_{k,j}^\dagger \hat a_{k,j},
$}}
where $k$ labels waveguide modes and $j$ labels the port number. 
Details on the derivation of $H_{\text{Ring}}$ and the waveguide coupling are given in supplementary material S.1.

\begin{figure}[t]
	\begin{center}
		\includegraphics[width=.95\columnwidth]{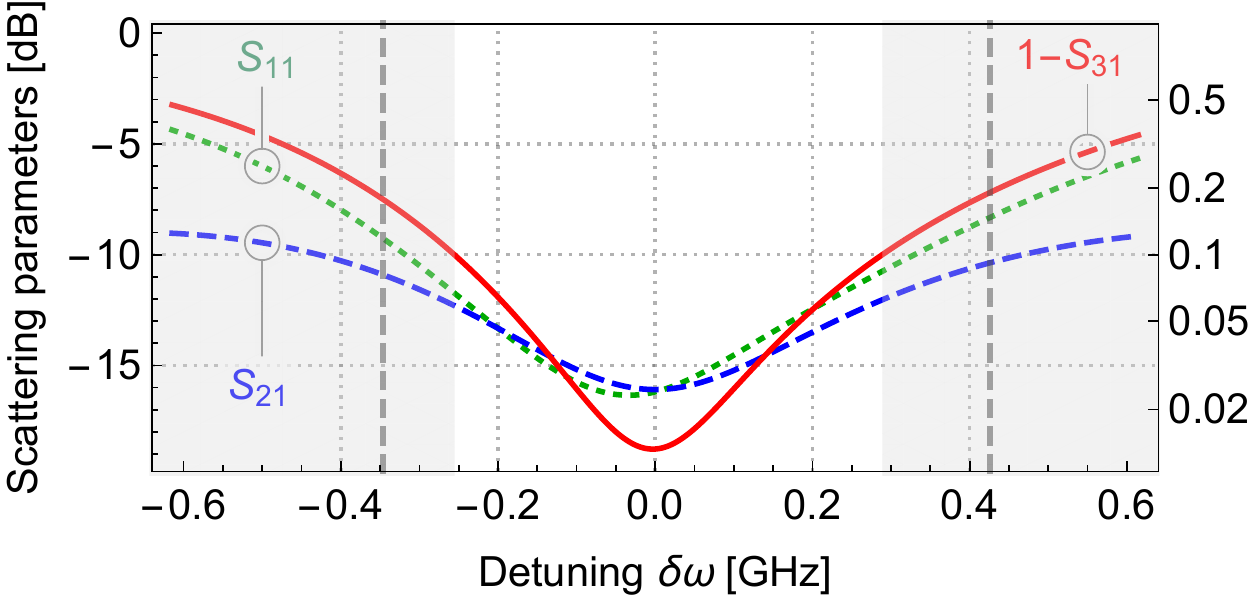}
		\caption{Scattering parameters in dB as function of detuning from the optimal signal frequency $\delta\omega = \omega_{k} - \omega_{\text{\text{opt}}}$. 
			Vertical dashed lines indicate the position of excited states of the ring, which get transiently excited in the scattering process (see 
			supplementary material S.4.)
			The bandwidth at the -10 dB point is $>500 $~MHz, as indicated by the grey shaded areas.
		}
		\label{fig:BW}
	\end{center}
\end{figure}

\paragraph*{Scattering calculations:} For the purposes of this Letter, we assume coherent field inputs at each port and we quantify the scattering of waveguide modes from the ring structure 
using the SLH formalism~\cite{Combes:2016, Muller:2017} which allows us to calculate output field amplitudes and photon fluxes.  
We note that the SLH formalism can be adapted to non-classical input fields~\cite{Baragiola:PRA:2012, Fan:PRB:2014, Baragiola:PRA:2017}. 
Assuming single mode input, the time-evolution of the density matrix for the open ring, $\rho$, satisfies the master equation
\begin{align}
	\dot \rho = -\ii \comm{H_{\text{SLH}}}{\rho} + {\sum}_{j} \diss{b_{j} } \rho,
	\label{eq:SLHFull}
\end{align}
with $H_{\text{SLH}} = H_{\text{Ring}} + H_{\text{D}}$ and
\begin{align}
	H_{\text{D}} &= -\tfrac{\ii}{2} {\sum}_{j} g_{k} \big( \alpha_{j} \ee^{-\ii \omega_{k} t} q^{(j   )}_{+} + \HC \big) \,,\\
	b_{j} &= g_{k}\: q^{(j)}_{-} + \alpha_{j} \ee^{-\ii \omega_{k} t} \mathds1 \,,
\end{align}
where $\alpha_{j}$ is the field amplitude of the incoming signal in transmission line $j$ at frequency $\omega_{k}$.
The frequency-dependent coupling strengths $g_{k}$ are calculated in supplementary material S.1
and depend on the physical realisation.  
The outgoing field amplitudes $\beta$ and photon fluxes $B$ into port $j$ are then given by
\begin{align}
	\beta_{j}=\Tr\{b_{j} \rho\}, \quad
	B_{j}=\Tr\{ b_{j}\hc b_{j} \rho\} \,.
\end{align}

To calculate scattering dynamics, we first diagonalise $H_{\text{Ring}}$ in a truncated Hilbert space of the  dynamical variables, $\hat n'_{1,2}$. 
Retaining eigenmodes with eigenumbers $ n'_{1,2}=-4,-3,...,4$ is sufficient to accurately describe low energy ring modes, $\{\ket{E_0},\ket{E_1},\ket{E_2},...\}$.
We then further truncate the ring Hilbert space to the lowest $l$ modes for scattering calculations. 
Typically, $l=3$ to $5$ is sufficient for calculating scattering matrices, due to the strongly anharmonic spectrum of the ring.  
{Apart from controlled truncations, we do not make secular or other approximations in $H_{\text{Ring}}$.}

We characterise   {circulation} using the steady-state photon-flux scattering matrix $\mathds{S}$ which relates the input and output photon fluxes in each port,
\mbox{$ \mathbf{B} = \mathds{S} \mathbf{A} $},
with the vector of input photon fluxes \mbox{$\mathbf{A} = \big\{ \!\abss{\alpha_{1}}, \abss{\alpha_{2}}, \abss{\alpha_{3}}\!\big\}$}.
The scattering matrix elements are given by \mbox{${S}_{ij} = \lim_{t\rightarrow\infty}B_{i}(t) / \abss{\alpha_{j}} \geq 0$}. 
For an ideal, passive, three-port, clockwise circulator
\begin{align}
	\mathds S = \l( \begin{array}{ccc}
			0 & 1 & 0 \\
			0 & 0 & 1 \\
			1 & 0 & 0
		\end{array}\r) .
\end{align}

\begin{figure}[t]
	\begin{center}
		\includegraphics[width=.95\columnwidth]{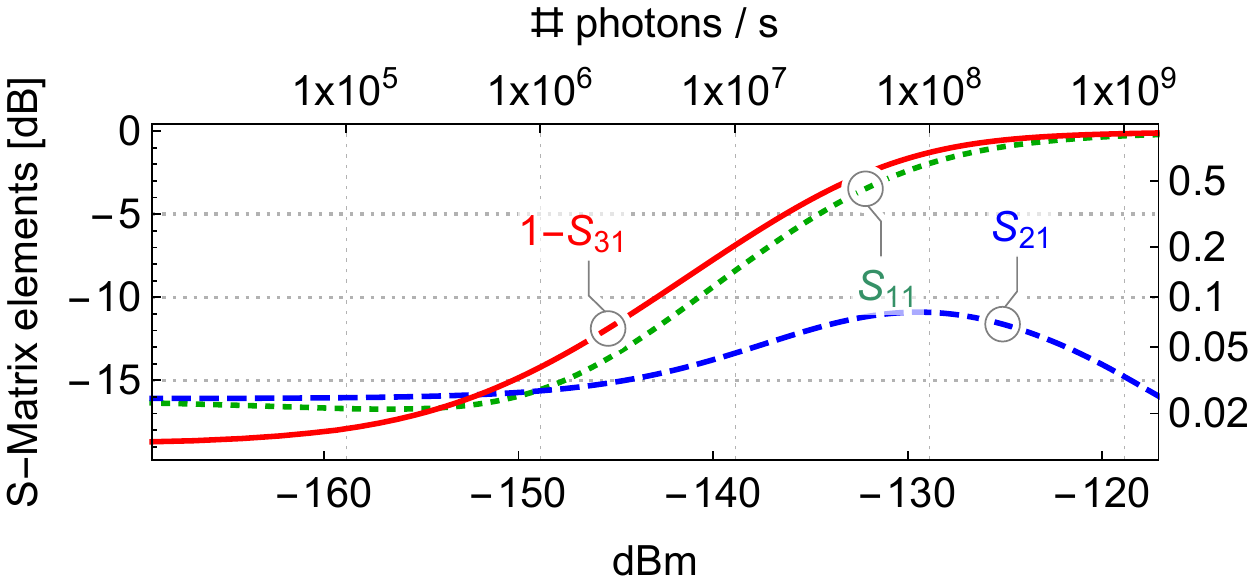}
		\caption{
			S-matrix elements (in dB) as function of input power at the optimal signal frequency and central bias, demonstrating the strong non-linearity of the ring.
		}
		\label{fig:NonLin}
	\end{center}
\end{figure}

The scattering of photons from  the ring is mediated by excitations of the ring modes. Notionally, the central bias is tuned 
so that the relevant ring modes are anti-degenerate with respect to the signal frequency $\omega_{k}$, i.e.\ for any ring mode with eigenenergy $\omega_{k} + E_r$, 
there is a dual mode with eigenenergy $\omega_{k}-E_r$.  
At this tuning point, the scattering matrix becomes maximally non-reciprocal, leading to perfect circulation. 
Here, we choose the bias point so that circulation proceeds clockwise. 
We note that there is another bias point where the circulation is reversed, see supplementary material S.3.

Here, we choose circuit parameters that are feasible for both QPS and JJ implementations, as detailed in supplementary material S.3.
The tunnelling energy is $E_{T} / \hbar = 15$~GHz and the kinetic energy term $E_{\Sigma}/\hbar = 7.80$~GHz.
The segments are biased equally with $\Nn^{(k)} = N_{\mathrm S, \text{opt}}= 1/3$, such that the conserved charge of the ground-state is $N_{0} = 1$.
For this choice, we find perfect clockwise circulation at a central bias of $X_{\text{opt}} = 0.356$ and with an input signal frequency of $\omega_{\text{opt}} = 12.293$~GHz.
At this optimal point, the coupling strength to the waveguides is $g_{\text{opt}} = 1.832$~GHz.  
{This value is large, but experimentally feasible~ \cite{FornDiaz:NP:2017, Yoshihara:NP:2017}.  
Since $g_{\text{opt}} \ll \omega_{\text{opt}}$, the rotating wave approximation implicit in the SLH formalism is still reasonable.
The circulation characteristics at different input frequencies $\omega_{\text{in}}$ can be found by simple scaling of all energies in the problem by the desired ratio $\omega_{\text{in}} / \omega_{\text{opt}}$.
}
 
\paragraph*{Bandwidth and nonlinearity:} \fig{fig:BW} shows the spectral response of the ring to a weak coherent field incident on port 1, 
with ideal operating parameters for clockwise circulation. 
We achieve an insertion loss approaching $-20$dB with reflection and isolation both below $-15$dB.
The performance of the circulator degrades to $-10$dB at detunings of $\sim\pm250$ MHz, so that the $-10$dB bandwidth exceeds \mbox{500 MHz}.

The ring structure is realised as a coherent, non-linear superconducting device, so it has an anharmonic spectrum and will saturate at sufficiently high powers.  
Since we diagonalise $H_{\text{Ring}}$ non-perturbatively, the  SLH formalism enables us to quantify the non-linear response of the system~\cite{Combes:2016} {to continuous incident fields}.
\fig{fig:NonLin} shows the normalised output flux of the circulator versus input powers.  
The 1dB compression point, where performance degrades by 1dB relative to the ideal, linear case is at $-156$dBm, corresponding to $\sim 10^{5}$ incident photons per second. 
{Useful circulation extends much further in power, with  10dB of isolation at $-140$dBm.}

{For transient incident fields, e.g.\ for a Fock-state with some specific temporal envelope, 
we expect that dispersion within the transmission window will induce envelope distortion, analogous to Refs.~\cite{Fan:PRB:2014,Sathya:PRL:2014}.
}

\begin{figure}[t]
	\begin{center}
		\includegraphics[width=.95\columnwidth]{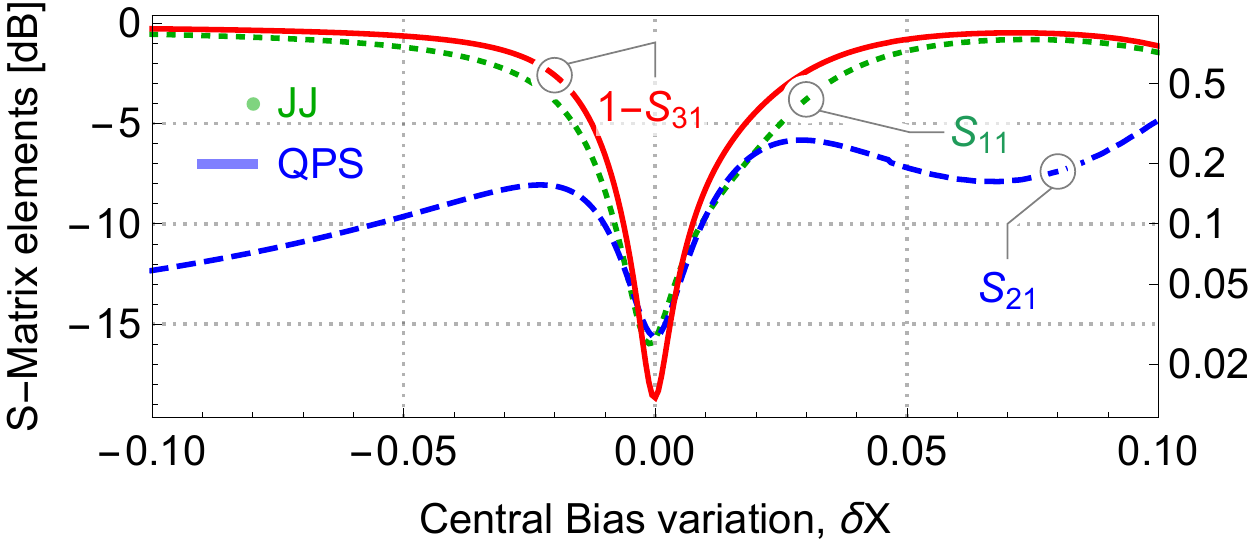}
		\caption{
			S-matrix elements  as a function of the variation in the central bias parameter, $\delta\Px = \Px - \Px_{\text{opt}}$ (in dB).  
			For the JJ ring, $\Px=\Phi_X/\Phi_0$; for the QPS ring, $\Px=Q_X/ (2e)$.  
			The notional output is to port 3; outputs to other ports arise from bias variations.  
			The bars in the left of the figure indicate typical variation in the central bias for a JJ-based implementation due to flux noise~\cite{Vion:S:2002},  
			and a QPS-based implementation due to charge noise~\cite{Koch:PRA:2007}.  
		}
		\label{fig:ChargeBW}
	\end{center}
\end{figure}

\paragraph*{Noise and disorder:} As in other circulator proposals, for ideal operation our scheme requires precise control of parameters. 
In our design this is the central bias, $\Px$, and the offset biases of each segment $\Nn^{(j)}$, 
as well as precise fabrication of the nodes, so that the tunnelling and mass terms are identical.  
In reality, all of these are subject to variation -- either  drifts in the bias parameters due to environmental noise, or fabrication imperfections, which are built into the device.

Analysing the effect of such imperfections, we focus on the quasi-static noise case, {where the dynamical time-scales of the noise processes are slower than the 
scattering.}
For the high-bandwidth devices we are considering here, and at moderately low input powers, 
the ring will stay predominantly in its ground-state at all times during operation {so that we can neglect 
decoherence due to non-radiative decay of ring states.
For the strong loss case, where non-radiative decay dominates over the coupling to the waveguides, circulation suffers as photons are lost to the environment. 
Since 
non-radiative rates of modern superconducting devices are much smaller than the timescale of circulation defined by the bandwidth, we neglect such processes in the following.}
Additionally, the eigenenergies of all ring-states for the parameters chosen are $>5$~GHz and thus well above usual operational temperatures of superconducting quantum circuits 
$\sim 10$mK $\approx 200$MHz.  We now quantify the effect of quasi-static variations away from ideality for ring parameters. 

The central bias controls the degree of non-reciprocity through changes in the eigenstates of the ring.  
\fig{fig:ChargeBW} shows the variation of $S_{31}$ as the central bias is tuned away from the optimal point $X_{\text{opt}}$.  
Evidently, good circulation is maintained as long as fluctuations in the central bias are kept below $\sim1$\%.  
Also shown 
are bars (bottom left) indicating typical fluctuations in the central bias for the two implementations we consider.  
In this case, flux noise in the JJ implementation is exceedingly small, $\sim 10^{-4}\Phi_{0}$~\cite{Vion:S:2002}, 
whereas charge noise in the QPS implementation is at the $10^{-2} (2e)$ level~\cite{Koch:PRA:2007}.

\begin{figure}[!t!]
	\begin{center}
		\includegraphics[width=.9\columnwidth]{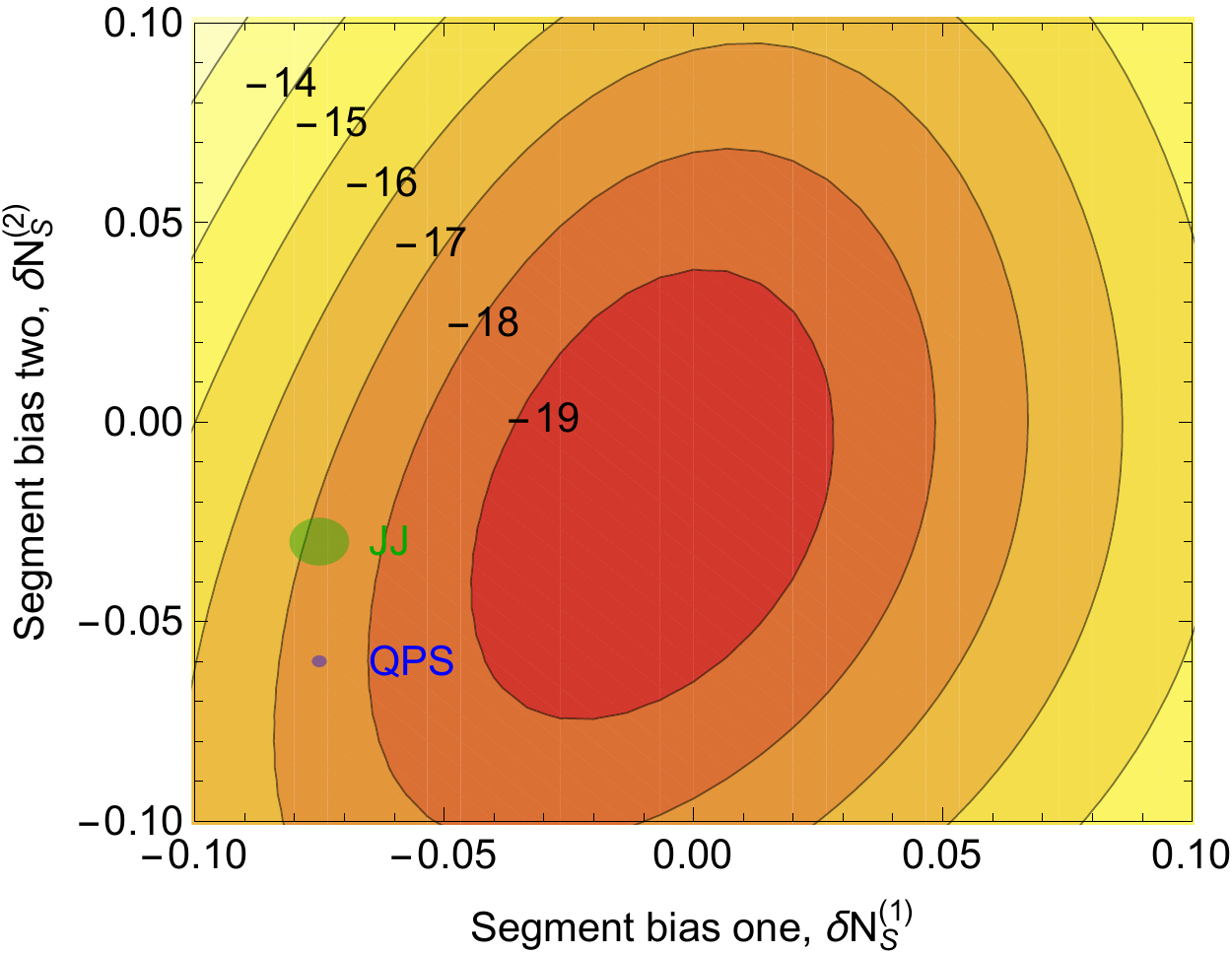}
		\caption{
			Effect of deviations in the node bias on the forward transmittance $1-S_{31}$ (in dB), with $\delta N_{S}^{(k)} = N_{S}^{(k)} - N_{S,\text{opt}}$.  
			The disks at the left of the diagram indicate the scale of typical slow bias fluctuations in JJ devices due to charge noise~\cite{Koch:PRA:2007} and QPS devices due to flux noise~\cite{Vion:S:2002}. 
		}
		\label{fig:nodebias}
	\end{center}
\end{figure}
\fig{fig:nodebias} shows contours of ${S}_{31}$ 
as two of three of the segment biases, $\Nn^{(1,2)}$, are varied by $\pm 0.1 p_{0}$.  
Typical scales for slow variations are shown in the error disks, bottom left.  
In this case, charge noise in the JJ implementation is at the $10^{-2}(2e)$ level, whereas flux bias noise in the QPS implementation is not visible on this scale. 
In either case, the system is relatively insensitive at the scale of these variations.  
We note that variation of the third segment bias, $\Nn^{(3)}$, 
is quantitatively similar to variations in $\Nn^{(2)}$.

\begin{figure}[!t!]
	\begin{center}
		\includegraphics[width=.9\columnwidth]{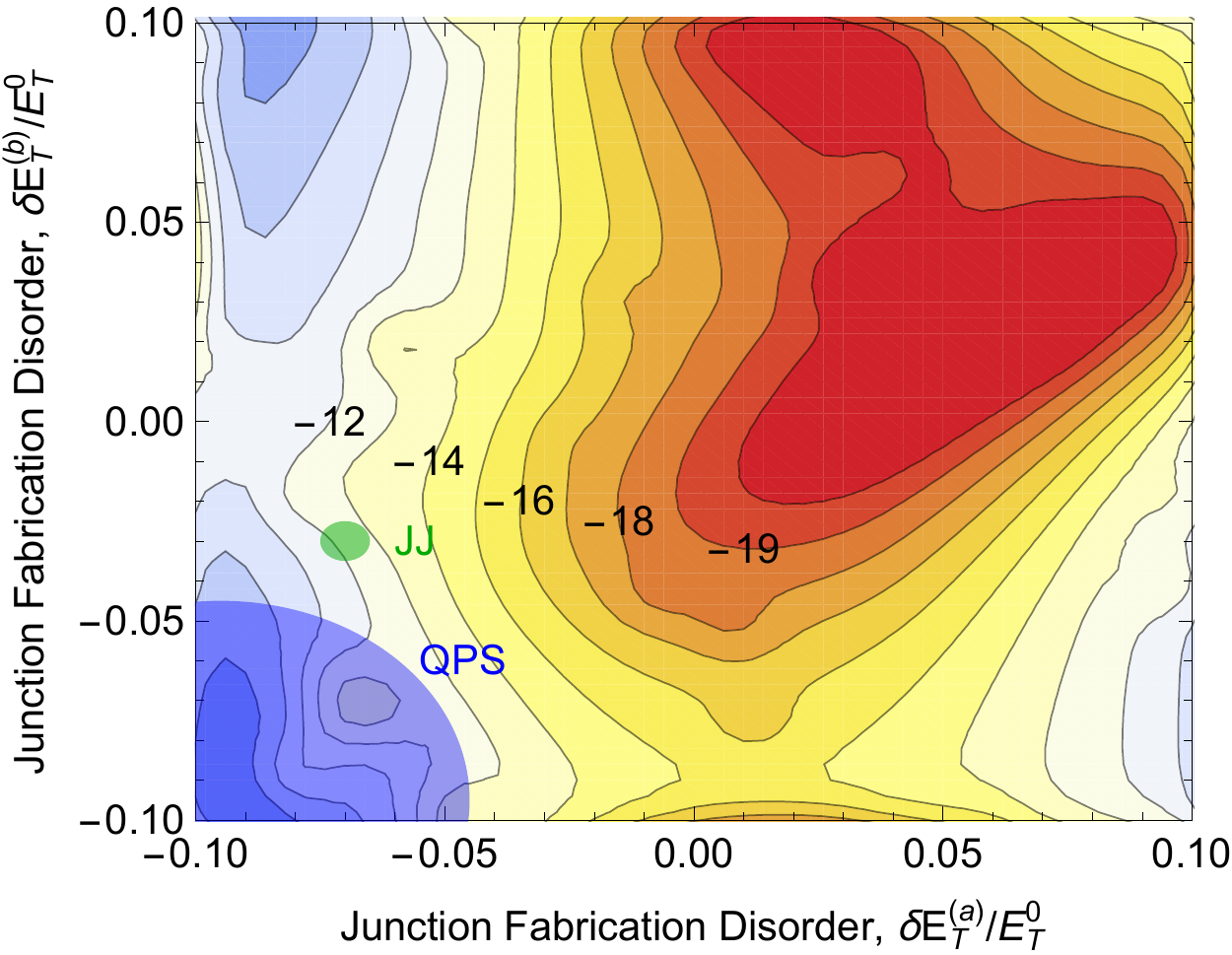}
		\caption{
			Effect of imperfections in the tunnelling energies $E_{T}^{(a,b)}$ on forward transmittance $1-S_{31}$ (in dB). 
			For each point in parameter space, central bias $X$ and segment biases $N_{S}^{(k)}$ 
			are optimised independently to find maximum forward transmission.
			The disks at the left indicate the scale of fabrication disorder in JJ~\cite{Fink:PRL:2009, Oliver:Private} and QPS devices~\cite{Peltonen:PRB:2013, Skacel:Private}.  
		}
		\label{fig:Disorder}
	\end{center}
\end{figure}
Lastly, \fig{fig:Disorder} shows the effect of fabrication imperfections on two of the  tunnelling energies, $E_T^{(a,b)}$.  
These are likely to be more variable than the mass terms, since the tunnelling is exponentially dependent on  device geometry. 
We note that when calculating scattering with imperfect tunnel junctions, {optimal circulation occurs at different bias points for each realisation of disorder}. 
Since fabrication disorder is static, this type of variation can be taken into account at initial tuneup of the devices, and each point in~\fig{fig:Disorder} 
represents an independent optimisation of the central bias $X$ and the segment biases $N_{S}^{(k)}$. 
Independent variations of the segment biases can partially counteract the asymmetry in junction parameters, evident from the large plateau in~\fig{fig:Disorder}.
Historically, more effort has been spent optimising JJ fabrication than QPS devices, so JJ parameters  are currently under better experimental control. 
At the bottom left,
disks indicate a typical scale for the reproducibility of the tunnelling energy for JJ, $\sim 1\%$, and QPS, $\sim 10\%$, 
when comparing junctions fabricated simultaneously on the same chip.
{We have also simulated the effect of disorder in other Hamiltonian parameters 
and find qualitatively similar results to Fig.~\ref{fig:nodebias},
as described in supplementary material S.5.}

In conclusion, we have shown that passive microwave circulators can feasibly be built from a ring of superconducting tunnel junction.  
The circuits can be integrated on chip with current fabrication technology and do not require any additional microwave or rf circuitry.  
The operating bandwidth is limited by the achievable waveguide coupling strength and can reach $>500$~MHz for reasonable parameters. 
Due to the anharmonic spectrum of the central ring structure, non-linearities are significant and the scattering matrix is strongly power dependent.
The dual implementations we propose are reasonably insensitive to disorder and noise in bias charges.  
At their current state of development, fabrication of QPS wires is less repeatable than JJ's, however there may be applications for each implementation. 

\acknowledgements{
	We thank A.~Blais, T.~Duty, S.~Filip, W.~Guichard, N.~Roch, H.~Rotzinger, P.~Scarlino and S.~Skacel for discussions and comments.
	This work was supported by the Australian Research Council under the Discovery and Centre of Excellence funding schemes (project numbers DP140100375, CE110001013 and CE170100039). 
	Computational resources were provided by the NCI National Facility systems at the Australian National University 
	through the National Computational Merit Allocation Scheme supported by the Australian Government.
}

\bibliography{../Circulator}

\newpage

\appendix

\renewcommand\thesection{S.\arabic{section}}
\renewcommand\thesubsection{\thesection.\Alph{subsection}}
\renewcommand\theequation{S-\arabic{equation}}
\renewcommand\thefigure{S.\arabic{figure}}
\renewcommand\thetable{S.\arabic{table}}

\section{Ring Hamiltonian \label{app:H}}

Here we derive the Hamiltonian of the circuits in Fig.~1
in the main text explicitly. 
We will focus on the QPS design, but point out differences and similarities to the dual JJ ring where important.
For the QPS geometry, the Hamiltonian of the central ring is~\cite{Koch:PRA:2010, Vool:2017}
\begin{align}
	H_{\text{QPS}} =& 1/2 (\hat{\mathbf{\phi}} - \mathbf{\phi}_{\mathrm{S}}) \mathds{L}^{-1} (\hat{\mathbf{\phi}} -\mathbf{\phi}_{\mathrm{S}}) \nn\\
		& - E_{S} \sum_{k} \cos{( \frac{2\pi }{2e}( q_{k+1} - q_{k} - Q_{x}/3) )} \,,
	\label{eq:HQPS}
\end{align}
with the inductance matrix and its inverse
\begin{align}
	\mathds{L} &= \l( \begin{array}{ccc} 
			L_{\Sigma} - L_{S} & -L_{S} & -L_{S} \\
			-L_{S} & L_{\Sigma} - L_{S} & -L_{S} \\
			-L_{S} & -L_{S} & L_{\Sigma} - L_{S} \\ 
		\end{array}\r) \,, \\
	\mathds{L}^{-1} &= \l( \begin{array}{ccc} 
			l_{1} & l_{2} & l_{2} \\
			l_{2} & l_{1} & l_{2} \\
			l_{2} & l_{2} & l_{1}\\ 
		\end{array}\r) \,,
\end{align}
with the effective ring inductance $L_{\Sigma} = 3L_{S} + L_{g} + L_{c}$ 
and the inverse inductances $l_{1} = L_{S} + L_{g} + L_{c} / L_{\Sigma}(L_{g} + L_{c})$ and $l_{2} = L_{S}  / L_{\Sigma}(L_{g} + L_{c})$. 
Note that $l_{1}-l_{2} = 1/L_{\Sigma}$.
Here we have assumed a perfectly symmetric ring with three equal QPS junctions, each with phase slip energy $E_{S}$ and intrinsic junction inductance $L_{S}$. Each ring segment couples to the outside transmission lines through the coupling inductances $L_{c}$ and has an additional parasitic inductance $L_{g}$, 
relevant for biasing of the segments.
The flux vector $\hat{\mathbf{ \phi}} = \l\{\hat{\phi}_{1}, \hat{\phi}_{2}, \hat{\phi}_{3}\r\}$ describes the flux degrees of freedom of each ring segment, with the conjugate charge variables $q_{k}$. The bias charge on the central island $Q_{\text x}$ is maintained through an external voltage, 
and each ring segment is pre-biased with an external magnetic flux $\phi_{\mathrm{S}}^{(k)}$, where $\mathbf{\phi}_{\mathrm S} = \l\{ \phi_{\mathrm{S}}^{(1)}, \phi_{\mathrm{S}}^{(2)}, \phi_{\mathrm{S}}^{(3)} \r\}$ is the vector of segment biases.

The JJ ring in Fig.~1
~(b) is described by the dual Hamiltonian to Eq.~\eqref{eq:HQPS}, where we exchange the role of charge and flux degrees of freedom and the inductance matrix becomes a capacitance matrix.

In the following, we replace flux operators by number operators, as $\phi_{k} = \Phi_{0} n_{k}$, with the the superconducting flux quantum $\Phi_{0} = h/2e$. 
Similarly, we can write $q_{k} = (2e) n_{k}$ in the JJ ring Hamiltonian, where charge of the Cooper pair takes the role of the superconducting flux quantum.

Then, performing a canonical transformation of the flux degrees of freedom in the Hamiltonian, as 
\begin{align}
	\l( \begin{array}{c} n_{1}' \\ n_{2}' \\ n_{3}' \end{array} \r) =
	\l( \begin{array}{ccc}  
		1 & 0 & 0 \\
		0 & -1 & 0 \\
		1 & 1 & 1 
	\end{array} \r) \l( \begin{array}{c} n_{1} \\ n_{2} \\ n_{3} \end{array} \r)
\end{align}
and corresponding on the canonical variables $q_{k}$ (see App.~\ref{app:canonical}), it becomes evident that the variable $n_{3}' = n_{1} + n_{2} + n_{3} = N_{0}$, corresponding to the total flux number of the ring, is conserved, i.e., $\partial H / \partial \phi'_{3} = 0$.
We can then rewrite the Hamiltonian as
\begin{align}
	\H_{\text{QPS}} = \frac{\Phi_{0}^{2}}{L_{\Sigma}} \Biggl\{ & \l( \hat{n}_{1}' - \frac12 (N_{0} + \Nn^{(1)} - \Nn^{(2)}) \r)^{2} \nn\\
		+& \l( \hat{n}_{2}'+ \frac12(N_{0} + \Nn^{(2)} - \Nn^{(3)}) \r)^{2} - \hat{n}_{1}' \hat{n}_{2}' \Biggr\} \nn\\
		- E_{S} \Biggl\{ &\cos{( \frac{2\pi }{2e}(q_{1}' - Q_{x}/3))} + \cos{( \frac{2\pi }{2e}(q_{2}' - Q_{x}/3))} \nn\\
		+& \cos{(\frac{2\pi }{2e}(q_{1}' + q_{2}' + Q_{x}/3))} \Biggr\} \,,
	\label{eq:HRing2}
\end{align}
and similar for the JJ ring. 
Here, $N_{0}$ is assumed constant and set by bias and parameter conditions and $\Nn^{(k)} = \mathbf{\phi}_{\mathrm{S}}^{(k)} / \Phi_{0}$. Specifically it is the initial flux biases through each ring segment which determine the constant total flux number in the ground-state.

\subsection{Coupling to external degrees of freedom}

The Hamiltonian describing the frequency dependent coupling between the QPS ring and a single mode of the external transmission lines is 
\begin{align}
	H_{\text{C,QPS}} = -\ii \Phi_{0} \sqrt{\frac{\hbar \omega_{k}}{2 L_{r}}} L_{c} \l( \hat{\mathbf{a}}_{k} - \hat{\mathbf{a}}\hc_{k} \r) \mathds L^{-1} ( \hat{\mathbf{n}} - \mathbf{N}_{\mathrm S} )
\end{align}
where 
we expressed the charge and phase variable of the transmission lines as
\begin{align}
	q_{k} 
		&= \sqrt{\frac{\hbar}{4Z}} \l( a_{k} + a\hc_{k} \r) \,,\nn\\
	\phi_{k} 
		&= -\ii \sqrt{\frac{\hbar Z}{4}} \l( a_{k} - a\hc_{k} \r) \,,
\end{align}
and $\omega_{k}$ is the eigenfrequency of the mode at which we probe the coupling. $Z = \sqrt{L_{r}/ C_{r}}$ is the transmission line impedance.

In the JJ case, the coupling Hamiltonian is 
\begin{align}
	H_{\text{C,JJ}} 
		&=2e \sqrt{\frac{\hbar \omega_{k}}{8 C_{r}}} C_{c} \l( \hat{\mathbf{a}}_{k} + \hat{\mathbf{a}}\hc_{k} \r) \mathds C^{-1} ( \hat{\mathbf{n}} - \mathbf{N}_{\mathrm S} ) \,,
\end{align}
i.e., we couple to the opposite quadrature of the waveguide field than in the QPS case. 

After the canonical transformation, and taking advantage of the conserved total flux number $n_{3}' = N_{0}$, we write the interaction as
\begin{align}
	H_{\text{C,QPS}} = \sum_{k}g_{k} \Bigl\{ & \hat{a}_{k,1} \l( \hat{n}_{1}' + \Nn^{(1)'} \r) 
		+ \hat{a}_{k,2} \l( -\hat{n}_{2}' + \Nn^{(2)'} \r) \nn\\
		+ & \hat{a}_{k,3} \l( -\hat{n}_{1}' + \hat{n}_{2}' + \Nn^{(3)'}) \r)
		+ \text{h.c.} \Bigr\} \,,
	\label{eq:HCoup}
\end{align}
with the same expression valid for the JJ ring circulator. 
Here the effective bias values for each of the ring segments are
\begin{align}
	N^{(1)'}_{n} &= L_{\Sigma} ( l_{2} (N_{0} - \Nn^{(2)} - \Nn^{(3)} ) - l_{1} \Nn^{(1)} ) \,,\nn\\
	N^{(2)'}_{n} &= L_{\Sigma} ( l_{2} (N_{0} - \Nn^{(1)} - \Nn^{(3)} ) - l_{1} \Nn^{(2)} ) \,,\nn\\
	N^{(3)'}_{n} &= L_{\Sigma} ( l_{1} (N_{0} - \Nn^{(3)} ) - l_{2} (\Nn^{(1)} + \Nn^{(2)}) ) \,.
\end{align}
Above we have also defined the frequency dependent coupling constants $g_{k} / \hbar = -\ii \Phi_{0} L_{c}/L_{\Sigma}\sqrt{\hbar \omega_{k} / 2 L_{r} }  $ for the QPS-ring 
and $g_{k} / \hbar = (2e) C_{c}/C_{\Sigma} \sqrt{\hbar \omega_{k} / 2C_{r} } $ for the JJ-circulator.

%

\section{Parameters\label{App:Para}}

Since we require a well-defined Aharonov-Casher/Bohm phase to realise non-reciprocity, we need the device to operate in a parameter regime where the flux/charge-number of each ring segment is a well-defined quantum number. 
To this end, we choose the ratio of potential energy and kinetic energy $E_{T} / E_{\Sigma} \approx 2$.

Table~\ref{tab:Para} details the parameters we have used in our numerical simulations for QPS and JJ ring devices~\cite{Skacel:Private}. 
They have been chosen such that they are well within reach of current fabrication technology for both physical implementations.
Numbers are chosen such that both implementations have the same energy scales, so that they behave equivalently in the simulations.
\begin{table}[ht!]
	\begin{tabular}{ c|c|c|c } 
		QPS-ring & & JJ-ring & \\
		\hline
		$E_S/\hbar$ & $15$GHz & $E_J$ & $15$GHz \\ 
		\hline
		$L_g$ & $2400$nH & $C_g$ & $57.63$fF  \\
		\hline
		$L_c$ & $100$nH & $C_c$ & $2.40$fF  \\
		\hline
		$L_s$ & $900$nH & $C_J$ & $21.61$fF \\
		\hline
		$E_{\Sigma}/\hbar = \Phi_{0}^{2} / L_{\Sigma} \hbar$ & $7.80$GHz & $E_{\Sigma}/\hbar = (2e)^{2} / C_{\Sigma} \hbar$ & $7.80$GHz \\
	\end{tabular}
	\caption{Parameters used for numerical simulations. 
		Numbers are chosen equivalently for both QPS and JJ ring, where inductances and capacitances are related through $L \rightarrow C = (2e)^{2} / \Phi_{0}^{2} L$.
	}
	\label{tab:Para}
\end{table}

\section{Bias parameters for circulation\label{app:BiasIdeal}}

\begin{figure}[t]
	\begin{center}
		\includegraphics[width=\columnwidth]{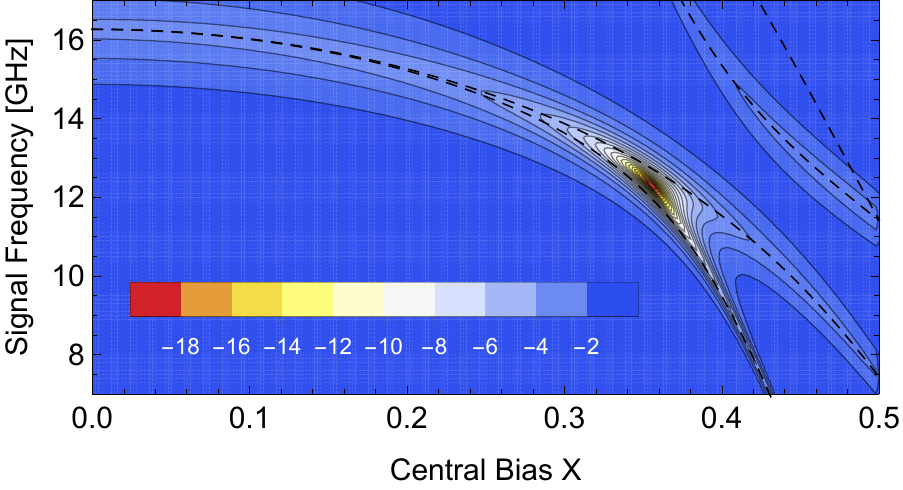}
		\caption{Normalized power transmittance $1-S_{31}$ in dB as function of signal frequency $\omega_{k}$ and applied bias charge $\Px$. 
			Dashed lines indicated the energies of eigenstates of the ring structure. 
		}
		\label{fig:biasContour}
	\end{center}
\end{figure}

In order to find the parameters for perfect circulation, we systematically vary the signal frequency $\omega_{D}$ and central bias $X$ of the ring 
and calculate the power scattering matrix $\mathds{S}$ for an input field incident on port one. Fig.~\ref{fig:biasContour} shows the scattering parameter $S_{31}$ as function of signal frequency and central bias. 
Maximum clockwise circulation with the parameters detailed below is achieved for an optimal input signal frequency of $\omega_{\text{opt}} = 12.293$~GHz at a central bias of $X_{\text{opt}}=0.356$.
We note that further increasing the central bias to values $X>0.5$ will invert the direction of circulation. The structure of the eigenstates of the ring is mirror-symmetric with respect to $X=0.5$, 
meaning that at central bias of $X=0.644$ we find counter-clockwise circulation.

\begin{figure}[htbp]
	\begin{center}
		\includegraphics[width=.95\columnwidth]{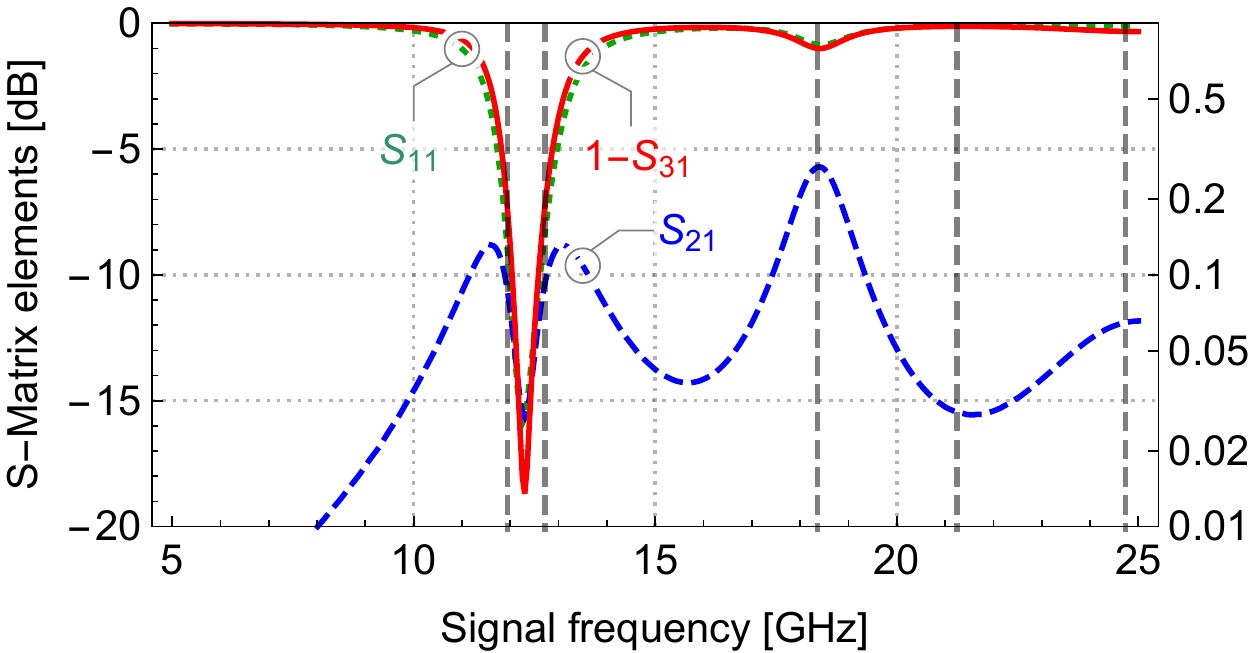}
		\caption{Scattering parameters in dB as function of signal frequency over a large range of input frequencies. Vertical dashed lines indicate the position of levels in the ring. 
			Maximum circulation is achieved when the signal frequency is between two adjacent levels which are both transiently excited, analogous to the classical Faraday circulator.
		}
		\label{fig:BWLarge}
	\end{center}
\end{figure}

Fig.~\ref{fig:BWLarge} shows the dependence of the scattering parameters on the signal frequency at the optimal central bias point $X_{\text{opt}}$. 
Optimal circulation is achieved when the drive frequency is chosen in between two of the ring levels, such that coupling into both levels is possible and interference takes place.

\section{Scattering calculations in input-output theory \label{app:NoRes}}

\subsection{Time-dependent scattering \label{app:TimeDep}}

\begin{figure}[htbp]
	\begin{center}
		\includegraphics[width=.95\columnwidth]{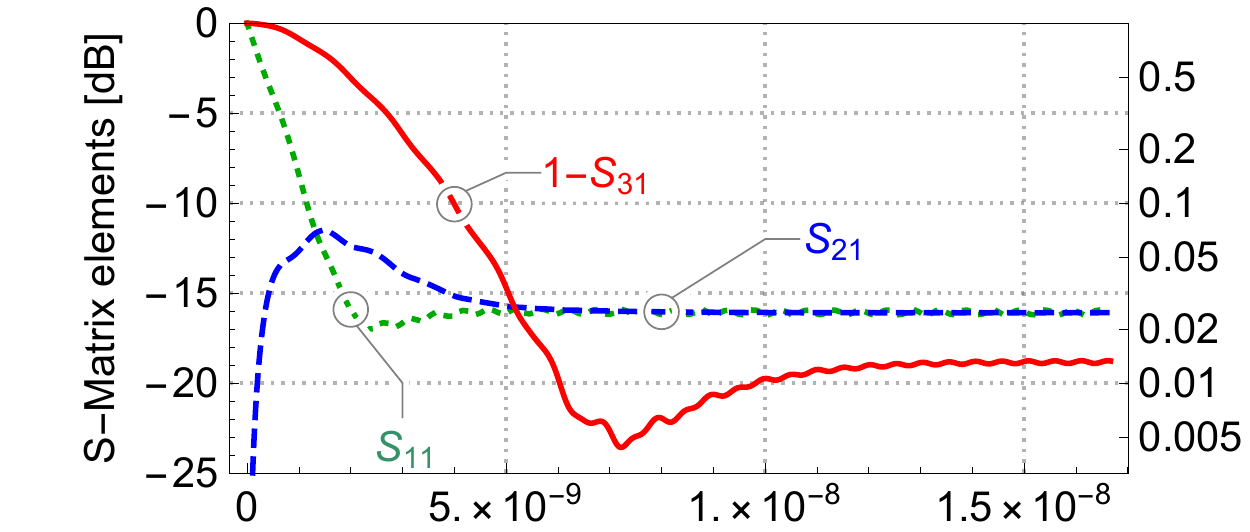}
		\includegraphics[width=.95\columnwidth]{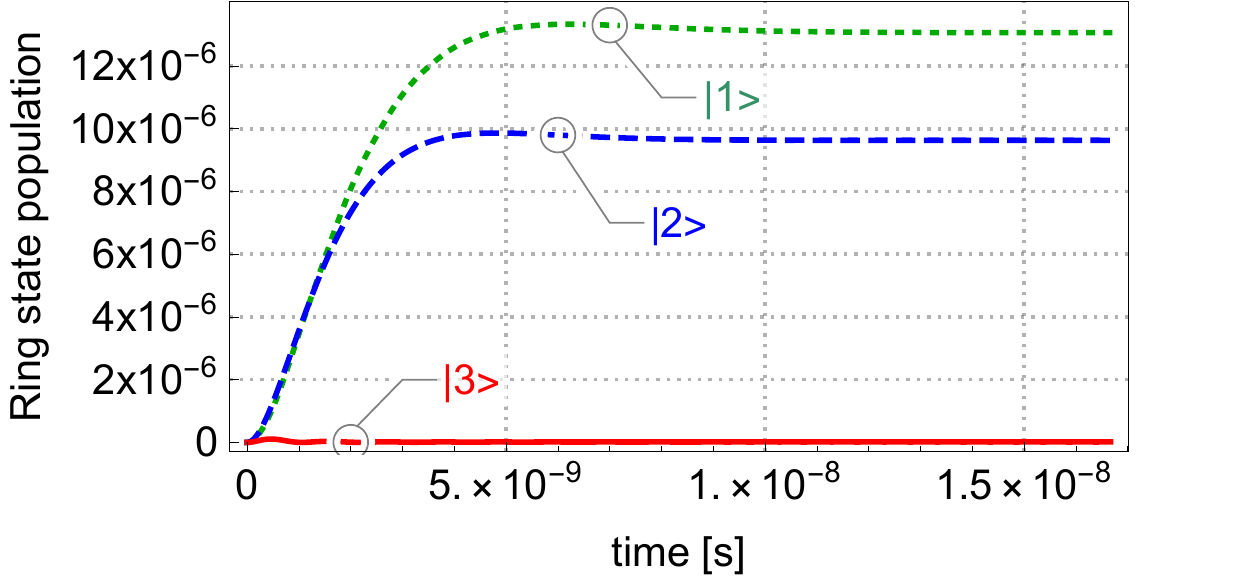}
		\caption{
			(Top) Transmittance to the three output ports as function of time at optimal driving frequency and bias. 
			(Bottom) Population of the ring states as function of time for the same parameters, demonstrating that only the first two excited states of the ring play a role in the scattering dynamics.
		}
		\label{fig:TimeNoRes}
	\end{center}
\end{figure}

The scattering problem defined in Eq.~(4) 
in the main text
is in general time-dependent through the input field. 
When calculating the scattering parameters we numerically solve the time-dependent master equation and evolve the system to its time-dependent steady-state, 
when any spurious dynamics due to the initial  ringup has completely decayed.
The results reported in this manuscript are then calculated from this long-time final state. 
As an example of the time-evolution of the system and scattering parameters, we show in Fig.~\ref{fig:TimeNoRes} the time evolution of the scattering parameters 
as well as of the population of the states of the ring. 
The final times in those plots is the final time taken in our simulations for these parameters.

\subsection{Time-independent steady-state calculations}

At or close to the ideal parameter conditions for circulation, when the drive frequency is located between two levels of the ring, the system can be well described by a time-independent master equation.
This is achieved by truncating the ring to the lowest three energy levels, and then moving into a rotating frame at the signal frequency. 
The drive Hamiltonian in Eq.~(4)
then becomes time-independent 
and the master equation can be solved for the steady-state of the system plus external drives. 
However, this approximation is only valid in a very narrow parameter range, where the drive frequency is close to resonant with several levels of the ring at the same time. This method is therefore unsuited for finding good operational parameters, 
but may facilitate faster and more efficient calculations once they are determined.

\section{Additional disorder calculations \label{App:Dissorder}}
{
For completeness we here show how circulation is affected by a spread in additional Hamiltonian parameters, apart from central bias, the node biases 
and the junction tunnelling energies already discussed in the main text. 
As such disorder will be a static effect arising from imperfections in fabrications, here for each realisation of disorder we have optimised the central bias value independently, 
as would be done at initial tune-up of these devices. 
}
\begin{figure}[!t!]
	\begin{center}
		\includegraphics[width=1\columnwidth]{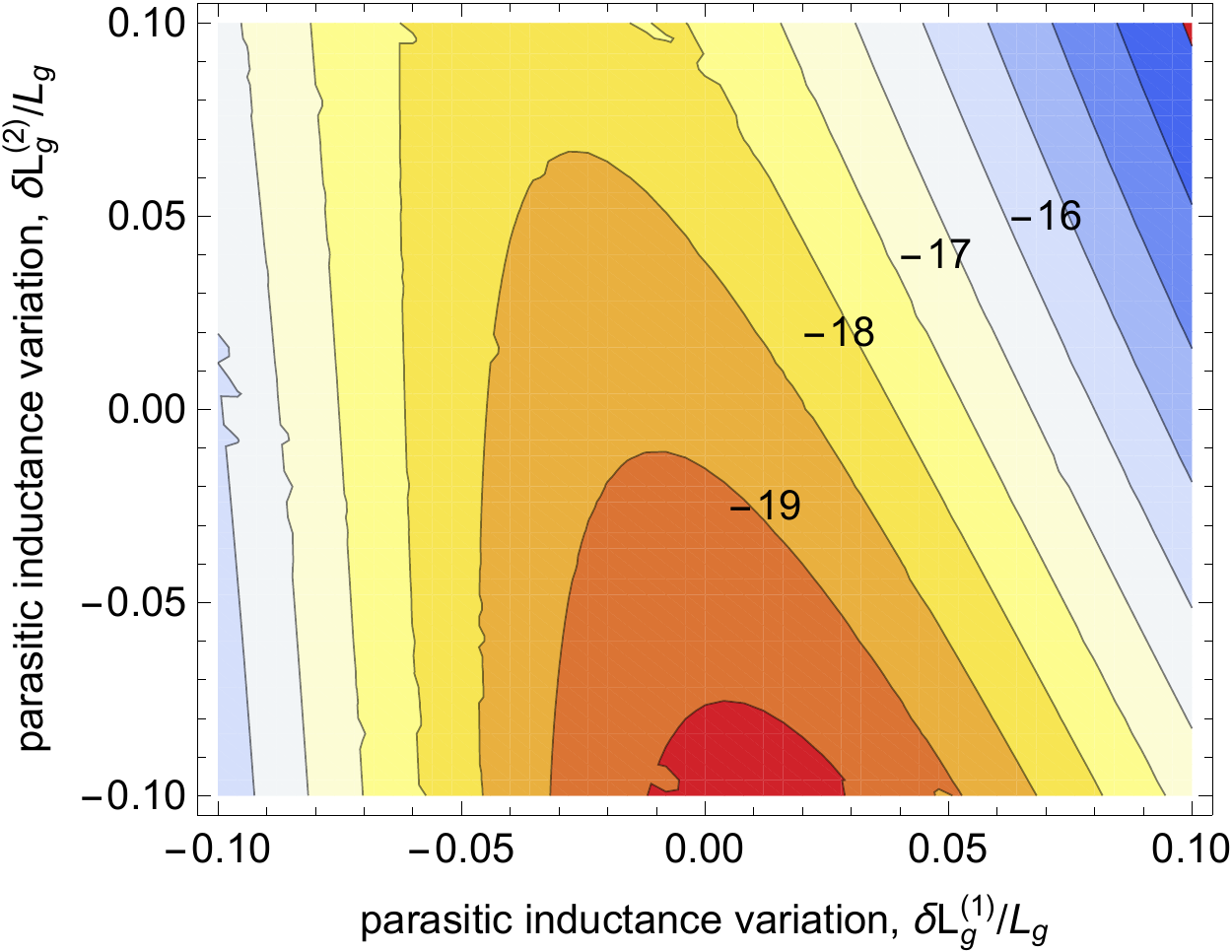}
		\caption{
			Effect of disorder in the parasitic inductances $L_{g}$ by $\pm10\%$ on the power circulated in clockwise direction, represented by the scattering parameter $1-S_{31}$. 
			Calculations are at the ideal point determined from optimising the central bias $X$.
			Other parameters as before, same picture holds in the JJ ring for variations in the parasitic capacitance $C_{g}$.
		}
		\label{fig:Cg}
	\end{center}
\end{figure}
{Fig.~\ref{fig:Cg} shows the result of variations in two of the parasitic inductances $L_{g}$. The asymmetry in the plot is due to the structure becoming asymmetric in these circumstances.}
\begin{figure}[!t!]
	\begin{center}
		\includegraphics[width=1\columnwidth]{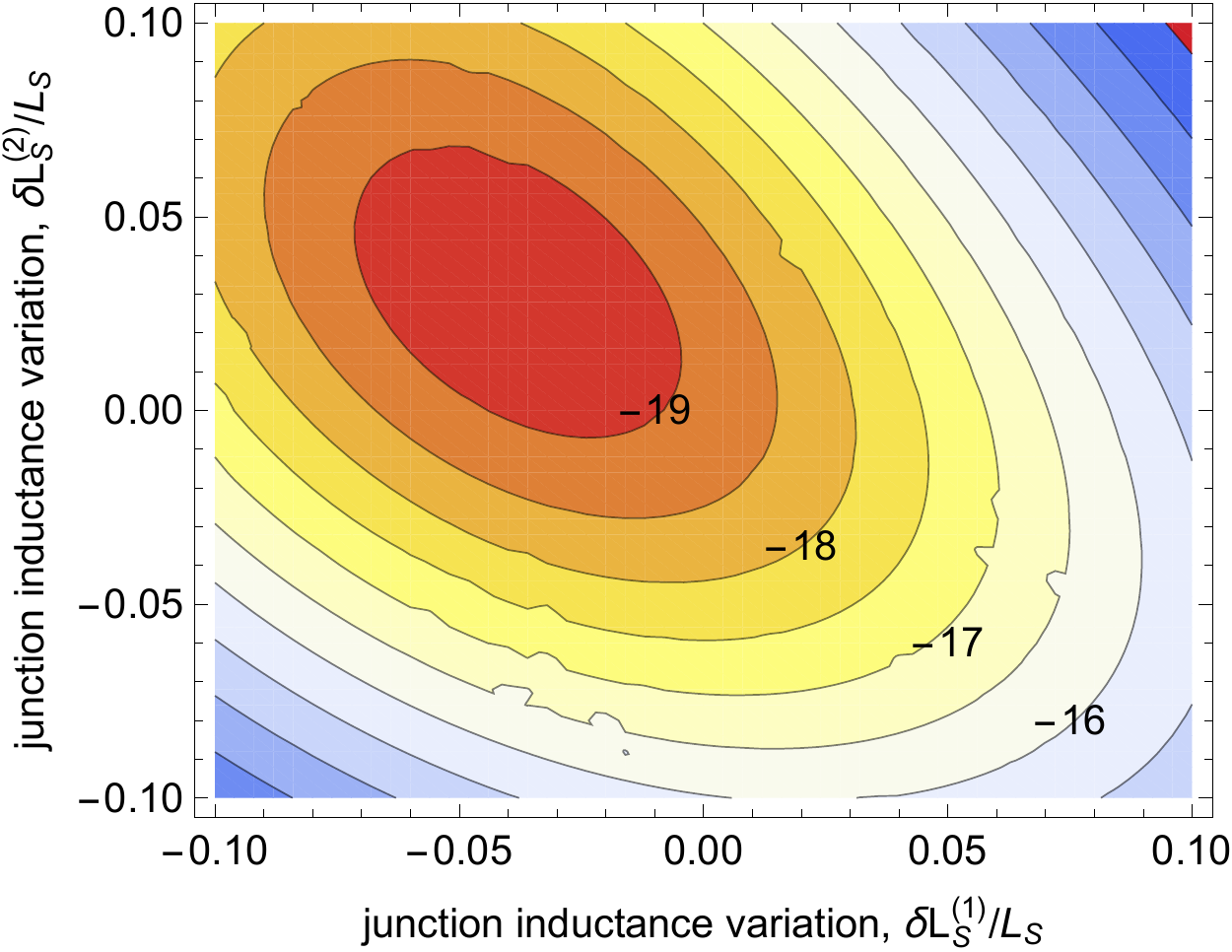}
		\caption{
			Effect of disorder in the junction inductances $L_{j}$ by $\pm10\%$.
		}
		\label{fig:Cj}
	\end{center}
\end{figure}
{Fig.~\ref{fig:Cj} shows the result of variations in two of the junction inductances $L_{s}$}.

\section{Canonical transformations for multiple variables\label{app:canonical}}

In order for a variable transformation to be considered canonical, the Euler-Lagrange equations of motion have to remain invariant. This implies that when we replace $q \rightarrow Q$, the canonical momentum associated with $q$ transforms as $p\rightarrow P = \frac{\partial \mathcal L}{\partial Q}$. 
Further, one can show that for each pair of variables this implies
\begin{align}
	\frac{\partial Q_{m}}{\partial q_{n}} = \frac{\partial p_{n}}{\partial P_{m}} \,.
\end{align}
Then it follows that if a vector of variables $\vec q$ transforms linearly, as $\vec q \rightarrow \vec Q$ with
\begin{align}
	\vec Q = \mathds A \vec q \,,
\end{align}
then the canonical momenta have to transform as $\vec p \rightarrow \vec P$ with
\begin{align}
	\vec P = \mathds B^{-1} \vec p
\end{align}
with the transformation matrix $\mathds B = \mathds A^{T}$.

\end{document}